\documentclass{pasj00}

\begin{document}
\SetRunningHead{T. Matsumoto}{An implicit scheme for ohmic dissipation
  with AMR}
\Received{2010/9/4}
\Accepted{2010/11/29}
\Published{2011/4/25}

\title{An Implicit Scheme for Ohmic Dissipation with Adaptive Mesh Refinement}

\author{Tomoaki \textsc{Matsumoto}}
\affil{Faculty of Humanity and Environment, Hosei University,
  Chiyoda-ku, Tokyo 102-8160, Japan}
\email{matsu@hosei.ac.jp}

%


\maketitle

\begin{abstract}
  An implicit method for the ohmic dissipation is proposed.  The
  proposed method is based on the Crank--Nicolson method and exhibits 
  second-order accuracy in time and space. The proposed method has been
  implemented in the SFUMATO adaptive mesh refinement (AMR) code.  The
  multigrid method on the grids of the AMR hierarchy converges the
  solution. The convergence is fast but depends on the time
  step, resolution, and resistivity. Test problems demonstrated that
  decent solutions are obtained even at the interface between fine and
  coarse grids.  Moreover, the solution obtained by the proposed method
  shows good agreement with that obtained by the explicit method,
  which required many time steps.  The present method reduces the
  number of time steps, and hence the computational costs, as compared with
  the explicit method.
\end{abstract}

\section{Introduction}

The magnetic field plays an important role in star formation. Taking
the magnetic field into account, simulations of protostellar
collapse have been performed in numerous studies (reviewed by
\cite{Klein07}).  Most of these studies assumed ideal magnetohydrodynamics (MHD).

Interstellar gas is partially ionized, and there are several processes
of magnetic diffusion, e.g., the ohmic dissipation, the Hall effect, and
ambipolar diffusion.  The ohmic dissipation is effective at high
densities of $n \gtrsim 10^{16}\,\mathrm{cm}^{-3}$, whereas the ambipolar
diffusion is effective at low densities of $n \lesssim
10^9\,\mathrm{cm}^{-3}$ (e.g., \cite{Kunz04}).  The timescale of the
magnetic diffusion is significantly longer than the freefall time at
$n \lesssim 10^{12}\,\mathrm{cm}^{-3}$ \citep{Nakano02}, and hence
magnetic diffusion does not appear to change the behavior of the
gravitational collapse qualitatively. The gravitational collapse
ceases in the dense region of $n \gtrsim 10^{11}\,\mathrm{cm}^{-3}$
owing to the formation of an adiabatic core, i.e., the first core \citep{Larson69}.
Therefore, subsequently formed objects, e.g., circumstellar disks,
protostars, and outflows, likely suffer from magnetic diffusion.

The recent numerical simulations for protostellar collapse begin to
take into account the magnetic diffusion (e.g., \cite{Machida06,Machida07}).
However, the governing equation of the magnetic diffusion is
parabolic, and therefore the time step for the magnetic diffusion is
very small compared with the hydrodynamic time step when a high-resolution explicit method is employed.  High resolution is important in the simulation of protostellar
collapse and is usually provided by means of adaptive mesh
refinement (AMR).

Several strategies for solving the magnetic diffusion have been
proposed.  The super time-stepping method is a type of explicit
method, in which a large time step can be used 
\citep{O'Sullivan06,O'Sullivan07,Choi09}.  However, for the diffusion
dominated problem, the time step is still restricted to be shorter than that
of the hydrodynamic time step.  \citet{Tilley08} proposed a 
semi-implicit scheme for ambipolar diffusion using a two-fluid
approximation, where the time step is restricted in inverse 
proportion to the drift velocity.  The present author previously
implemented the ohmic dissipation in a nested grid code by using a
sub-cycle of the induction equation \citep{Machida06,Machida07}. By
this method, the protostellar collapse from a molecular cloud core to protostar formation was successfully simulated.  Although
each sub-cycle required a small computational cost, the number of
sub-cycles becomes very large when solving the magnetically dissipative
region, e.g., the region proximal to and inside of a protostar.  Moreover, the
resistivity was approximated as being locally constant.

An implicit scheme for solving the ohmic dissipation has been
developed and implemented in the SFUMATO MHD-AMR code \citep{Matsumoto07}.
In \S2, the details of the implicit scheme are presented. In \S3, the
results of several numerical tests are presented. Finally, the paper is
summarized in \S4.

\section{Implicit scheme}
The induction equation with the ohmic dissipation is given by
\begin{equation}
\frac{\partial \mbox{\boldmath$B$}}{\partial t} =
\nabla \times \left( \mbox{\boldmath$v$}\times\mbox{\boldmath$B$} 
- \eta \nabla\times \mbox{\boldmath$B$}
\right),
\label{eq:induction}
\end{equation}
where $\mbox{\boldmath$B$}$, $\mbox{\boldmath$v$}$, and $\eta$ denote the magnetic field,
velocity, and resistivity, respectively.
Equation~(\ref{eq:induction}) is solved by an operator splitting
approach.  The contribution of the first term on the right-hand side of the equation
is solved explicitly according to \citet{Matsumoto07}, and the
contribution of the second term is then solved by the implicit scheme
presented herein.  We hereinafter restrict our focus to the solution of 
the ohmic dissipation.

\subsection{Discretization}
The governing equation of the ohmic dissipation is given by
\begin{equation}
\frac{\partial \mbox{\boldmath$B$}}{\partial t} = -\nabla\times\left(\eta\nabla\times\mbox{\boldmath$B$}\right).
\label{eq:basic}
\end{equation}
Equation~(\ref{eq:basic}) is written in conservation form as follows:
\begin{equation}
\frac{\partial \mbox{\boldmath$B$}}{\partial t} + \nabla \cdot \mbox{\boldmath$F$} = 0,
\label{eq:conservation}
\end{equation}
where the numerical flux $\mbox{\boldmath$F$} = (\mbox{\boldmath$F$}_x, \mbox{\boldmath$F$}_y, \mbox{\boldmath$F$}_z)$ is given by
\begin{equation}
\mbox{\boldmath$F$}_x = \eta
\left(
\begin{array}{c}
0 \\
-\partial_x B_y + \partial_y B_x \\
-\partial_x B_z + \partial_z B_x 
\end{array}
\right),
\label{eq:Fx}
\end{equation}
\begin{equation}
\mbox{\boldmath$F$}_y = \eta
\left(
\begin{array}{c}
-\partial_y B_x + \partial_x B_y \\
0 \\
-\partial_y B_z + \partial_z B_y 
\end{array}
\right),
\label{eq:Fy}
\end{equation}
\begin{equation}
\mbox{\boldmath$F$}_z = \eta
\left(
\begin{array}{c}
-\partial_z B_x + \partial_x B_z \\
-\partial_z B_y + \partial_y B_z \\ 
0 
\end{array}
\right).
\label{eq:Fz}
\end{equation}
Equation~(\ref{eq:conservation}) is discretized as follows:
\begin{equation}
\mbox{\boldmath$B$}_{i,j,k}-\mbox{\boldmath$b$}_{i,j,k} 
+\lambda \Delta t \left(\nabla \cdot \mbox{\boldmath$F$}\right)_{i,j,k}
+(1-\lambda)\Delta t \left(\nabla \cdot \mbox{\boldmath$f$}\right)_{i,j,k} =0,
\label{eq:difference}
\end{equation}
where, for convenience, the unknown variables are written in uppercase, and
the known variables are written in lowercase:
$\mbox{\boldmath$B$}:=\mbox{\boldmath$B$}^{n+1}$,
$\mbox{\boldmath$F$}:=\mbox{\boldmath$F$}^{n+1}$, 
$\mbox{\boldmath$b$}:=\mbox{\boldmath$B$}^{n}$, and 
$\mbox{\boldmath$f$}:=\mbox{\boldmath$F$}^{n}$. 
The superscript $n$ denotes the time level, and $\Delta t =
t^{n+1}-t^n$.  
The subscripts $i,j,k$ are the indexes of a cell in the $x$, $y$, and 
$z$ directions, respectively, and are used to label cells.
The parameter $\lambda$ specifies the type of temporal
difference. The backward difference is obtained when $\lambda=1$, and
the central difference is obtained when $\lambda=1/2$.  Therefore,
$\lambda=1$ results in a temporal first-order accuracy, while
$\lambda=1/2$ results in a temporal second-order accuracy.  The case of
$\lambda=1/2$ corresponds to the Crank--Nicolson scheme.  Spatial
discretization is performed with the central difference, yielding 
spatial second-order accuracy.  Each component of the numerical flux
is defined at the cell surface, and hence the divergence of the
numerical flux is calculated as follows:
\begin{eqnarray}
\left(\nabla \cdot \mbox{\boldmath$F$}\right)_{i,j,k}
&=& \frac{\mbox{\boldmath$F$}_{x,i+1/2,j,k}-\mbox{\boldmath$F$}_{x,i-1/2,j,k}}{\Delta x} \nonumber \\
&+& \frac{\mbox{\boldmath$F$}_{y,i,j+1/2,k}-\mbox{\boldmath$F$}_{y,i,j-1/2,k}}{\Delta y} \nonumber \\
&+& \frac{\mbox{\boldmath$F$}_{z,i,j,k+1/2}-\mbox{\boldmath$F$}_{z,i,j,k-1/2}}{\Delta z},
\end{eqnarray}
and $\left(\nabla \cdot \mbox{\boldmath$f$}\right)_{i,j,k}$ is calculated in the
same manner.
The differential terms in $\mbox{\boldmath$F$}_{x,i+1/2,j,k}$ are given by
\begin{equation}
\left(\partial_x B_y\right)_{i+1/2,j,k}
 = \frac{B_{y,i+1,j,k}-B_{y,i,j,k}}{\Delta x},
\end{equation}
\begin{eqnarray}
\lefteqn{
\left(\partial_y B_x \right)_{i+1/2,j,k}
} \nonumber \\
&=& \frac{B_{x,i+1,j+1,k}+B_{x,i,j+1,k}
-B_{x,i+1,j-1,k}-B_{x,i,j-1,k}}{4 \Delta y}. 
\nonumber \\
&\mbox{}&
\end{eqnarray}
The resistivity $\eta$ at the cell surface is given
by arithmetic average, e.g.,
\begin{equation}
\eta_{i+1/2,j,k} = \frac{\eta_{i+1,j,k}+\eta_{i,j,k}}{2}.
\end{equation}

Equation~(\ref{eq:difference}) is rewritten in the form of a
difference equation as follows:
\begin{equation}
{\cal L} \mbox{\boldmath$B$}_{i,j,k} = \mbox{\boldmath$S$}_{i,j,k} ,
\label{eq:elliptical}
\end{equation}
where 
\begin{equation}
{\cal L} \mbox{\boldmath$B$}_{i,j,k}= \mbox{\boldmath$B$}_{i,j,k} + \lambda \Delta t
\left(\nabla\cdot\mbox{\boldmath$F$}\right)_{i,j,k} ,
\end{equation}
\begin{equation}
\mbox{\boldmath$S$}_{i,j,k} = \mbox{\boldmath$b$}_{i,j,k}
 - (1-\lambda) \Delta t \left(\nabla\cdot\mbox{\boldmath$f$}\right)_{i,j,k} .
\end{equation}
Equation~(\ref{eq:elliptical}) indicates that the unknown
$\mbox{\boldmath$B$}$ is solved by the linear operator ${\cal L}$ and the
source term $\mbox{\boldmath$S$}$, which is a function of the known $\mbox{\boldmath$b$}$.

\subsection{Multigrid method}

Equation~(\ref{eq:elliptical}) is solved by the multigrid method.
Here, the strategy of the multigrid method is the same as that of
\citet{Matsumoto07}, who solved the scalar PDE of the Poisson
equation, while the present method solves the vector PDE of
equation~(\ref{eq:basic}).  Therefore, all of the procedures of
\citet{Matsumoto07} are extended to those for vectors, and full-weight
prolongation and averaging restriction are performed for each
vector component. Since the smoothing procedure depends on the
equation to be solved, it is newly developed as shown in
\S~\ref{sec:smoothing}.

The multigrid method ${\cal L}_\mathrm{FMG}^{-1}$ solves $\mbox{\boldmath$B$}^\mathrm{new}$  when the initial estimation $\mbox{\boldmath$B$}^\mathrm{guess}$ and the source term $\mbox{\boldmath$S$}$ are
given as follows: 
\begin{equation}
\mbox{\boldmath$B$}^\mathrm{new} = {\cal L}_\mathrm{FMG}^{-1}(\mbox{\boldmath$B$}^\mathrm{guess}, \mbox{\boldmath$S$}).
\label{eq:multigrid}
\end{equation}
Since equation~(\ref{eq:elliptical}) is linear, we use the multigrid
method iteratively, as follows:
\begin{equation}
\mbox{\boldmath$R$} = \mbox{\boldmath$S$} - {\cal L}\mbox{\boldmath$B$}^\mathrm{guess}
\label{eq:mgiteration1}
\end{equation}
\begin{equation}
\mbox{\boldmath$B$}^\mathrm{new} = \mbox{\boldmath$B$}^\mathrm{guess} + {\cal L}_\mathrm{FMG}^{-1} (0, \mbox{\boldmath$R$})
\label{eq:mgiteration2}
\end{equation}
\begin{equation}
\mbox{\boldmath$B$}^\mathrm{guess}
\leftarrow 
\mbox{\boldmath$B$}^\mathrm{new} .
\label{eq:mgiteration3}
\end{equation}
This iterative utilization of the multigrid method reduces every
component of a residual, $\mbox{\boldmath$R$}$.

\begin{table*}
\begin{center}
  \caption{Numbers of iterations for the multigrid method 
    \label{table:mgparam}}
  \begin{tabular}{lccc}
    \hline
    \hline
    Schemes          & V-cycle & Pre-smoothing & Post-smoothing \\
    \hline
    FMG on AMR        &  4 & 4 & 4 \\
    MLAT-FAS on AMR   &  2 & 4 & 4 \\
    FMG on base grid  &  2 & 2 & 2 \\
    \hline
  \end{tabular}
\end{center}
\end{table*}

  The multigrid method given by equation (\ref{eq:multigrid}) consists
  of (1) the full multigrid (FMG) cycle on the AMR hierarchical grids,
  (2) the multilevel adaptive technique (MLAT) with the full approximation
  scheme (FAS) on these grids, and (3) an FMG-cycle on the base grid
  \citep{Matsumoto07}.  These schemes have parameters: the numbers of
  iterations for V-cycle, pre-smoothing, and post-smoothing procedures
  in each grid level.  These parameters adopted through this paper are
  shown in table~\ref{table:mgparam}.  These parameters affect a
  convergence speed of the multigrid method; small numbers of these
  iterations slow the convergence while the computational cost is reduced.

  As shown in section~\ref{sec:numerical_test}, several cycles of the
  multigrid method given by equations
  (\ref{eq:mgiteration1})--(\ref{eq:mgiteration3}) reduce the residual
  by more than an order of magnitude. 
  In the numerical tests, we performed 20 cycles of the multigrid
  method in order to estimate solutions converged enough.

\subsection{Smoothing}
\label{sec:smoothing}
As a smoothing operator, the red-black Gauss--Seidel iteration is adopted.  
When equation~(\ref{eq:elliptical}) is solved separately for $\mbox{\boldmath$B$}_{i,j,k}$
in each vector component, and $\mbox{\boldmath$B$}_{i,j,k}$ is replaced by
$\mbox{\boldmath$B$}^\mathrm{updated}_{i,j,k}$, we obtain the following relationship:
\begin{equation}
\mbox{\boldmath$B$}^\mathrm{updated}_{i,j,k} = \mbox{\boldmath$B$}_{i,j,k}+
\left(
\begin{array}{c}
R_{x,i,j,k} /(1+\alpha_x) \\
R_{y,i,j,k} /(1+\alpha_y) \\
R_{z,i,j,k} /(1+\alpha_z)
\end{array}
\right),
\label{eq:GS}
\end{equation}
where 
\begin{equation}
\mbox{\boldmath$R$}_{i,j,k} = 
\left(
\begin{array}{c}
R_{x,i,j,k}\\
R_{y,i,j,k}\\
R_{z,i,j,k}
\end{array}
\right)
= \mbox{\boldmath$S$}_{i,j,k} - {\cal L}\mbox{\boldmath$B$}_{i,j,k},
\label{eq:residual}
\end{equation}
\begin{eqnarray}
\alpha_x &=& \lambda \Delta t 
\left(\frac{\eta_{i,j-1/2,k}+\eta_{i,j+1/2,k}}{\Delta y^2}\right.\nonumber\\
&+&
\left.
\frac{\eta_{i,j,k-1/2}+\eta_{i,j,k+1/2}}{\Delta z^2}\right),
\end{eqnarray}
\begin{eqnarray}
\alpha_y &=& \lambda \Delta t 
\left(\frac{\eta_{i,j,k-1/2}+\eta_{i,j,k+1/2}}{\Delta z^2}\right.\nonumber\\
&+&
\left.
\frac{\eta_{i-1/2,j,k}+\eta_{i+1/2,j,k}}{\Delta x^2}\right),
\end{eqnarray}
\begin{eqnarray}
\alpha_z &=& \lambda \Delta t
\left(\frac{\eta_{i-1/2,j,k}+\eta_{i+1/2,j,k}}{\Delta x^2}\right.\nonumber\\
&+&
\left.
\frac{\eta_{i,j-1/2,k}+\eta_{i,j+1/2,k}}{\Delta y^2}\right).
\end{eqnarray}
Equation~(\ref{eq:GS}) gives the approximate solution of
$\mbox{\boldmath$B$}^\mathrm{updated}_{i,j,k}$ for a given initial guess of
$\mbox{\boldmath$B$}_{i,j,k}$.  We adopt equation~(\ref{eq:GS}) as a smoothing
operator.  A red-black ordering is adopted for sweeping the grid.

\subsection{Time step}
  The AMR code was equipped with two modes
  of time-marching: an adaptive and a synchronous time-step mode.  In
  the former mode, a coarser grid has a longer time step than a finer
  grid, and this mode is appropriate for non-self-gravitational gases
  because the system equations are hyperbolic.  In the latter mode,
  every grid-level has the same time step, and this mode is
  appropriate for self-gravitational gases because the Poisson
  equation is elliptic.  For a problem including the ohmic
  dissipation, the synchronous time-step is adopted because the
  equation~(\ref{eq:basic}) is parabolic.  

\section{Numerical tests}
\label{sec:numerical_test}
\subsection{Sinusoidal diffusion problem}
\begin{figure}
  \begin{center}
    \FigureFile(80mm,80mm){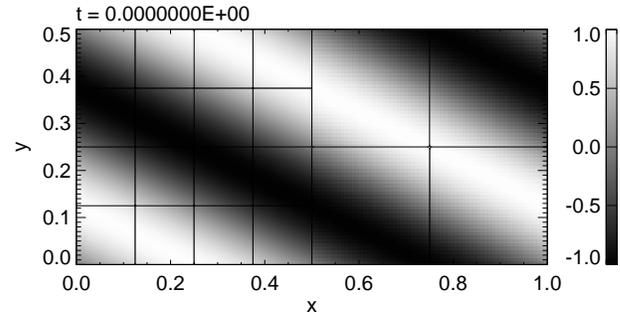}
  \end{center}
  \caption{Initial condition for the sinusoidal diffusion problem.
The gray scale denotes the distribution of $B_z$, and lines denote the
boundaries of the AMR blocks, each of which has $32^3$ cells.
}\label{amrAccuracyWave2ndInit.eps}
\end{figure}

We consider the problem in which the sinusoidal magnetic field diffuses with a constant resistivity.
The initial magnetic field is given as follows:
\begin{equation}
B_z = \sin(\mbox{\boldmath$k$} \cdot \mbox{\boldmath$r$}),
\end{equation}
and $B_x = B_y = 0$, where the wave number is set at $ \mbox{\boldmath$k$} =
2\pi (1,2, 0)^T$.  This setting reduces the equation of the ohmic
dissipation to the heat equation.  The resistivity is set at $\eta =
1$. The computational domain is $x, y, z \in [0, 1]\times[0,
1/2]\times[0, 1/4]$.  Periodic boundary conditions are imposed.  The
computational domain is covered by $4\times2\times1$ base blocks, each
of which has $N^3$ cubic cells.  The cell width is therefore given by
$\Delta x = \Delta y = \Delta z = 1/(4N)$ in the base grid.  The
domain of $x \in [0, 1/2]$ is refined by blocks that are twice as fine.
Figure~\ref{amrAccuracyWave2ndInit.eps} shows the initial distribution
of $B_z$ and the block distribution for $N=32$.  The cell width in the
left-hand side is $\Delta x = 1/256$, and that in the right-hand side is $\Delta x = 1/128$.

We performed the convergence test by changing the time step $\Delta
t$, in order to measure the temporal accuracy.  The $L_1$ norm of the
error is measured at $t = 4\times 10^{-3}$ through comparison with the exact
solution of the following equation: 
\begin{equation}
B_\mathrm{ex}(\mbox{\boldmath$r$}) = \exp\left(-\eta |\mbox{\boldmath$k$}|^2  t\right)
\sin(\mbox{\boldmath$k$} \cdot \mbox{\boldmath$r$}).
\end{equation}
The $L_1$ norm is estimated as follows:
\begin{equation}
L_1 = \frac{1}{V}\sum_{i,j,k} \left|B_z(\mbox{\boldmath$r$}_{i,j,k}) - B_\mathrm{ex}(\mbox{\boldmath$r$}_{i,j,k}) \right|
\Delta V_{i,j,k},
\end{equation}
where $\Delta V_{i,j,k}$ denotes the volume of a cell located at
$\mbox{\boldmath$r$}_{i,j,k}$, and $V$ denotes the volume of the entire
computational domain.  By the stage of $t = 4\times 10^{-3}$, the
amplitude of $B_z$ is reduced to 0.45.

\begin{figure}
  \begin{center}
    \FigureFile(80mm,80mm){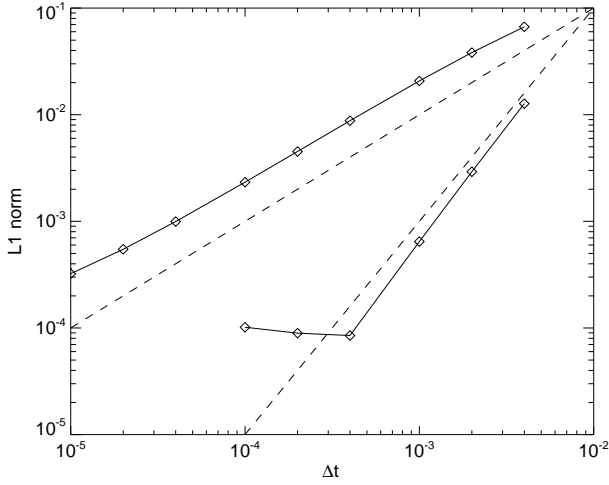}
  \end{center}
  \caption{
$L_1$ norm of error as a function of time step $\Delta t$ for
the sinusoidal diffusion problem.
The spatial resolution is $N=32$.
The upper and lower solid lines denote the errors for the cases in which 
$\lambda=1$ and $1/2$, respectively. The dashed lines indicate
the relationships of errors in proportion to $\Delta t$ and $\Delta t^2$, respectively.
}\label{amrAccuracyWave.eps}
\end{figure}

Figure~\ref{amrAccuracyWave.eps} shows the $L_1$ norm as a function of
the time step $\Delta t$.  We examined the cases of $\lambda=1$
(backward difference) and $\lambda=1/2$ (Crank--Nicolson).  The
scheme with $\lambda=1$ exhibits first-order accuracy, and that
with $\lambda=1/2$ exhibits second-order accuracy.  For the scheme
with $\lambda=1/2$, the decrease in the $L_1$ norm with decreasing $\Delta
t$ is saturated at $\Delta t \le 2\times10^{-4}$, exhibiting the constant
$L_1$ norm of $\sim 10^{-4}$.  The saturation is primarily attributed to
a discretization error.  We confirmed that the value of the saturation
decreases with decreasing cell width.

\begin{figure}
  \begin{center}
    \FigureFile(80mm,80mm){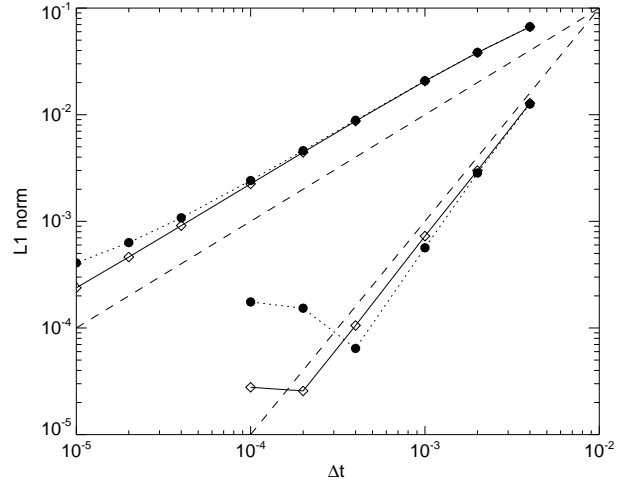}
  \end{center}
  \caption{
$L_1$ norm of error as a function of time step $\Delta t$ for the
sinusoidal diffusion problem.
The spatial resolution is $N=32$.
The upper and lower lines denote the errors for
the cases in which $\lambda=1$ and $1/2$, respectively.  
The solid lines with diamonds and 
the dotted lines with filled circles
denote the errors in 
$x \in [0, 1/2)$ (fine region) and 
$x \in (1/2, 1]$ (coarse region), respectively.
The dashed lines indicate
the relationships of errors in proportion to $\Delta t$ and $\Delta t^2$.
}\label{amrAccuracyWaveFC.eps}
\end{figure}

Figure~\ref{amrAccuracyWaveFC.eps} shows the $L_1$ norms estimated in
the fine region of $0 \le x < 1/2$ (fine region) and the coarse
region of $1/2 < x \le 1$ (coarse region).  For the
first-order scheme ($\lambda=1$), the error in the coarse grid is
slightly larger than that in the fine grid.  In contrast, the second-order scheme ($\lambda=1/2$) exhibits an error in the coarse grid that is approximately 6 times larger than that in the fine grid when $\Delta t \le 2\times10^{-4}$.

\begin{figure}[t]
  \begin{center}
    \FigureFile(80mm,80mm){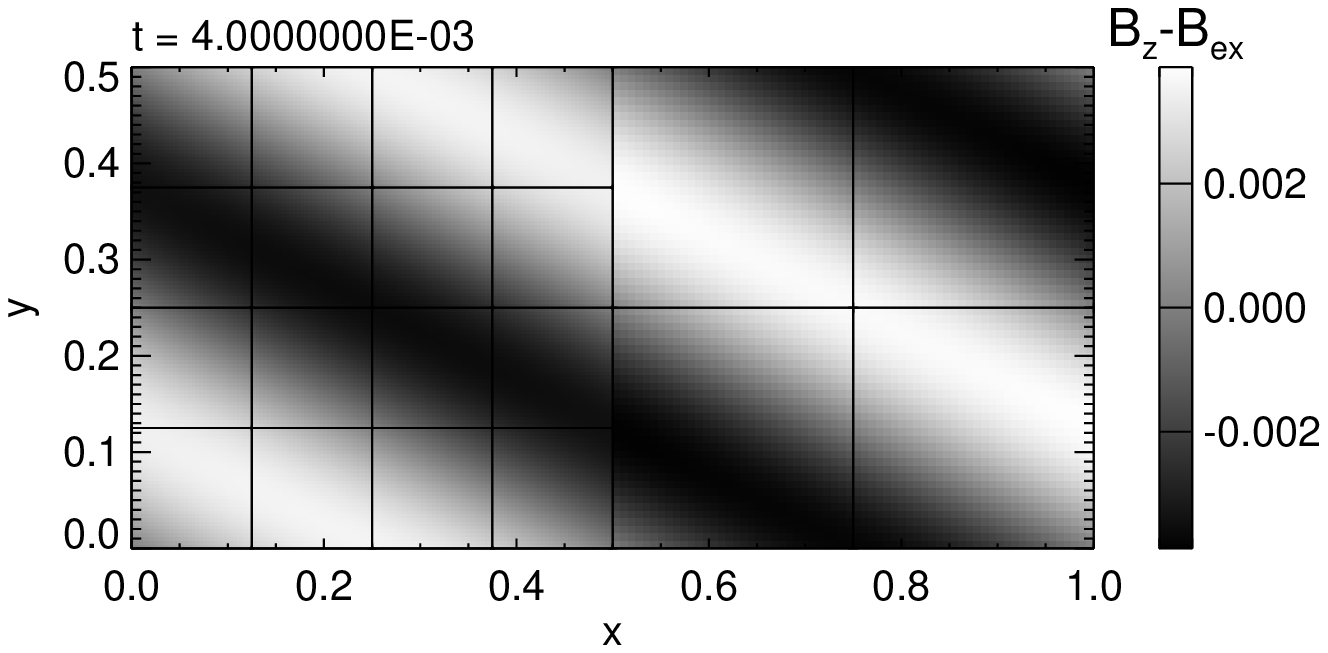}\\
    \FigureFile(80mm,80mm){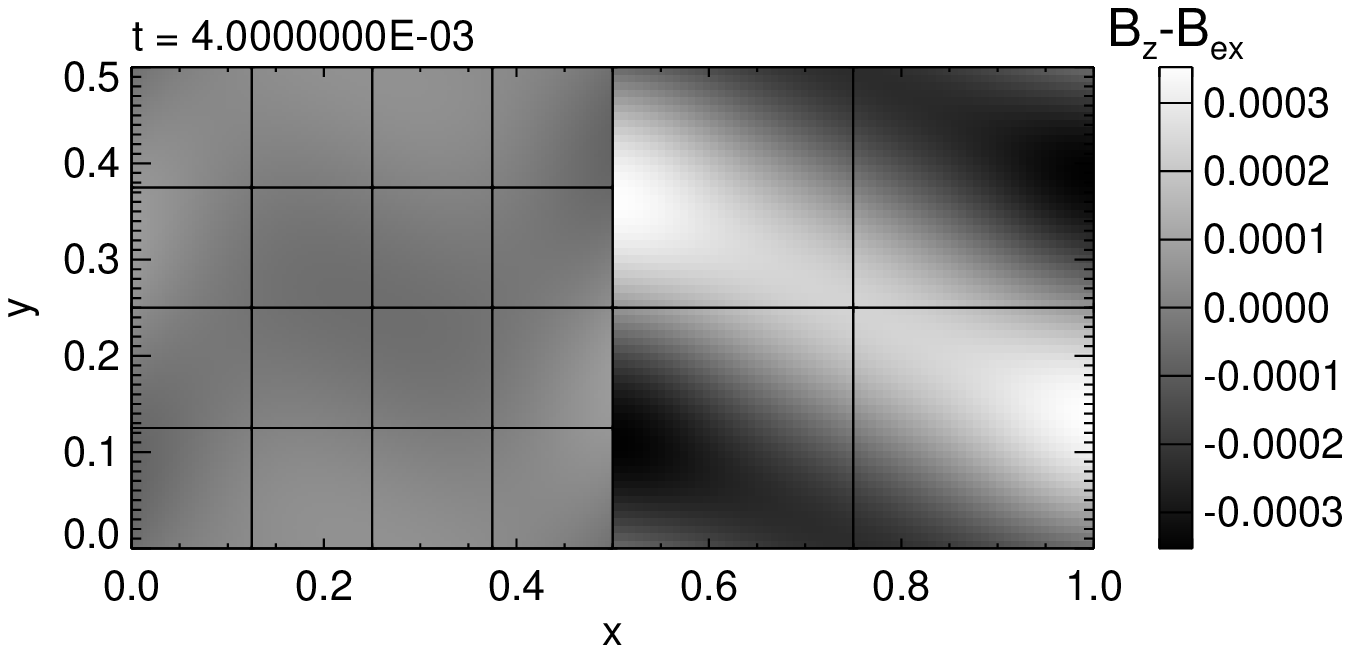}
  \end{center}
  \caption{ Distribution of error $B_z-B_\mathrm{ex}$ in the $x-y$
    plane for the sinusoidal diffusion problem.
    Errors are shown for 
    the $N=32$ resolution and $\Delta t = 10^{-4}$, solved by a scheme with $\lambda=1$ ({\it upper}) and a scheme that with $\lambda=1/2$ ({\it lower}).
  }\label{amr2DerrorWave.eps}
\end{figure}

Figure~\ref{amr2DerrorWave.eps} shows the distributions of the errors
at $\Delta t = 10^{-4}$ for the schemes with $\lambda=1$ and 1/2.
For the first-order scheme ($\lambda=1$), the error is distributed
smoothly through the fine and coarse grids.  For the second-order
scheme ($\lambda=1/2$), the coarse grid shows a larger 
systematic error than the fine grid, causing the large $L_1$ norm in the
coarse grid.  Moreover, the error is somewhat large in the coarse grid
near the interface between the fine and coarse grids.

\begin{figure}[t]
  \begin{center}
    \FigureFile(80mm,80mm){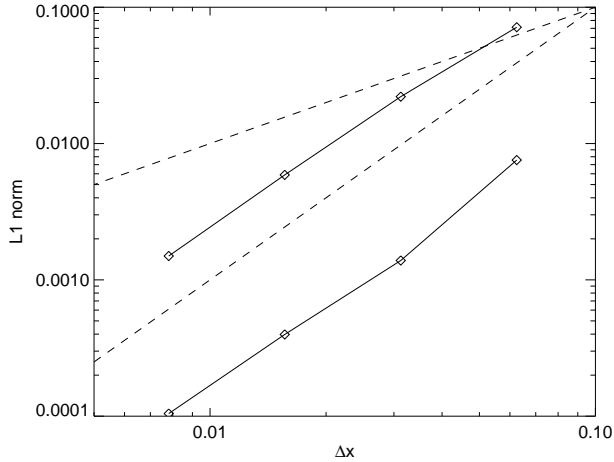}
\caption{
  $L_1$ norm of error as a function of cell width $\Delta x$ for the
  sinusoidal diffusion problem.
  Upper and lower solid lines indicate the errors obtained by the schemes with
  $\lambda=1$ and 1/2, respectively.
  The dashed lines indicate
  the relationships of errors in proportion to $\Delta t$ and $\Delta t^2$.
}
\label{amrAccuracyWaveSpace.eps}
\end{center}

We performed a convergence test with respect to the spatial
resolution by changing the number of cells inside a block, $N^3 = 4^3,
8^3, 16^3, 32^3$.  The time step is set to $\Delta t = 4\times10^{-3} (4/N)^2$.
Figure~\ref{amrAccuracyWaveSpace.eps} shows the $L_1$ norm as a
function of the cell width $\Delta x$ for the schemes with
$\lambda=1$ and 1/2.  Both schemes exhibit spatial second-order accuracy.

\end{figure}
\begin{figure}[t]
  \begin{center}
    \FigureFile(80mm,80mm){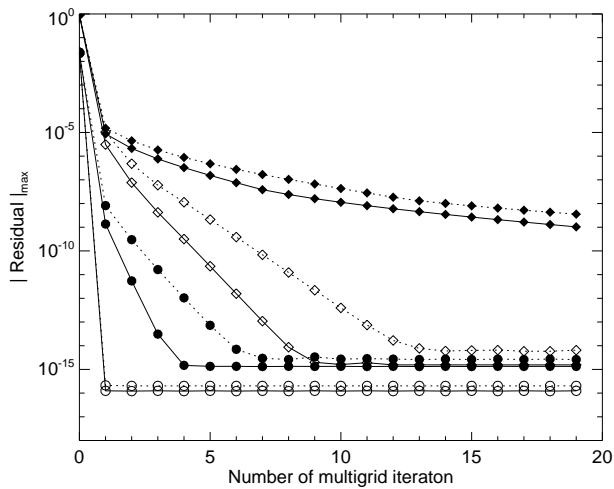}
\caption{
  Maximum residual $|\mbox{\boldmath$R$}|_\mathrm{max}$ as a function of iteration number of the multigrid method
  for the sinusoidal diffusion problem.
  Solid and dotted lines indicate data for $\lambda = 1/2$ and $1$, respectively. 
  Diamonds and circles indicate data for $\Delta t = 4\times10^{-3}$ and $10^{-4}$, respectively.
  Open and filled symbols indicate data for $N = 8$ and 32, respectively.
}
\label{res_ite.eps}
\end{center}
\end{figure}

Figure~\ref{res_ite.eps} shows the decrease in the maximum residual
during the iteration of the multigrid method given by
equations~(\ref{eq:mgiteration1}) through (\ref{eq:mgiteration3}) for 
various time steps $\Delta t$ and spatial resolutions $N$.  The
residual is calculated according to equation~(\ref{eq:residual}), and
$\max(|R_{x,i,j,k}|, |R_{y,i,j,k}|, |R_{z,i,j,k}|)$ is plotted.  In
all cases, one iteration of the multigrid method reduces the residual to
less than $10^{-5}$.  Comparison of the schemes with $\lambda=1/2$,
and 1 reveals that the scheme with $\lambda=1/2$ exhibits fast convergence.
Moreover, there is a tendency whereby cases with smaller $\eta \Delta t/
\Delta x^2$ exhibit faster convergence. The residuals with $\eta
\Delta t/ \Delta x^2 = 262, 16.4, 6.55$, and 0.410 in the fine grid
correspond to the lines with filled diamonds, open diamonds, filled
circles, and open circles, respectively.

\subsection{Gaussian diffusion problem}

\begin{figure}
  \begin{center}
    \FigureFile(80mm,80mm){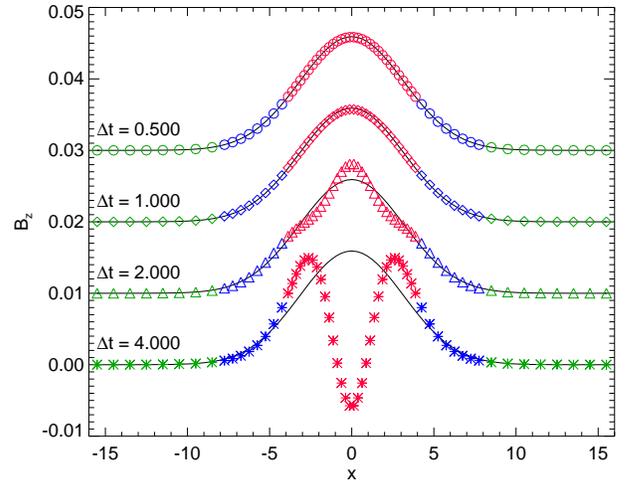}
    \caption{Distribution of $B_z$ in the $y=0$ plane at $t = 4$ for
      the Gaussian diffusion problem with $\lambda=1/2$.  Circles,
      diamonds, triangles, and asterisks denote the solutions of
      $\Delta t = 0.5$, 1.0, 2.0, and 4.0, respectively.  Green, blue, and
      red symbols indicate solutions on the grids of levels 0, 1, and 2,
      respectively.  The solid curve denotes the exact solution.  In order that all of the solutions could be plotted, the plots are offset from each other by 0.01 in the vertical
      direction.  }
    \label{odGaussianT1_2nd.eps}
  \end{center}
\end{figure}

\begin{figure}
  \begin{center}
    \FigureFile(80mm,80mm){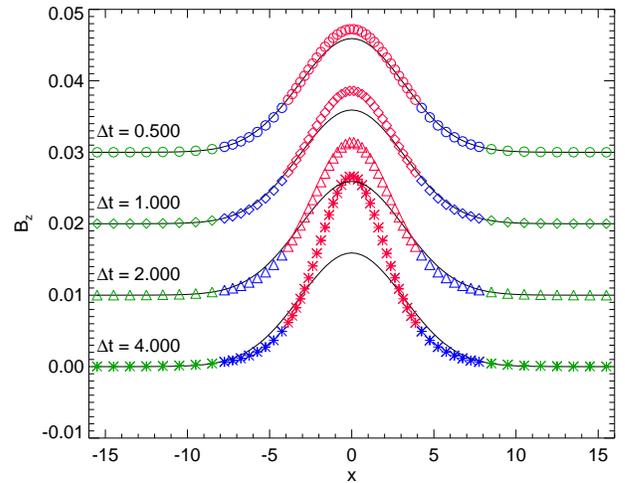}
    \caption{Same as Figure~\ref{odGaussianT1_2nd.eps}, but for $\lambda=1$.}
    \label{odGaussianT1_1st.eps}
  \end{center}
\end{figure}

We examine the diffusion of $B_z$ in the Gaussian profile, the exact
solution of which is given as follows:
\begin{equation}
B_z(x,y) = \frac{1}{4 \pi \eta (t+t_0)} \exp\left[-\frac{x^2+y^2}{4 \eta (t+t_0)}\right]
\end{equation}
where $t_0=1$ and $\eta=1$.  The computational domain is $x, y, z \in
[-16, 16]$, which is resolved by the base grid of $32^3$ cubic cells.
The region around the $z$ axis ($x=y=0$) is covered by the fine grids, as shown in
Figure~\ref{odGaussianT1_2nd.eps}.  The cell widths are $\Delta x =$
1.0, 0.5, and 0.25 for the grids of levels 0, 1, and 2, respectively. Periodic
boundary conditions are imposed.

Figure~\ref{odGaussianT1_2nd.eps} shows the solutions of $B_z$ at $t =
4$ with various time steps $\Delta t$ for the scheme of $\lambda=1/2$.
As the time step $\Delta t$ increases, the solution deviates from the
exact solution.  The solution with $\Delta t = 2.0$ shows significant
undulation $|x| \lesssim 4$. For the solution with $\Delta t = 4.0$,
the undulation mashes up the solution in the finest grid (red
asterisks).  Note that the Crank--Nicolson method is unconditionally
stable for the von Neumann stability analysis, while a large $\eta
\Delta t/\Delta x^2$ produces such undulation due to violation of the
maximum principle \citep{Morton05}.  We also found that the undulation
occurred with smaller $\Delta t$ when the initial Gaussian profile had
a narrower width (a smaller $t_0$).

The scheme with $\lambda=1$ yields smooth solutions even for large
$\Delta t$, as shown in Figure~\ref{odGaussianT1_1st.eps}.  Although
monotonicity is maintained in the solutions, the solution in the
finest grid deviated considerably from the exact solution when $\Delta
t$ is large.

\begin{figure}
  \begin{center}
    \FigureFile(80mm,80mm){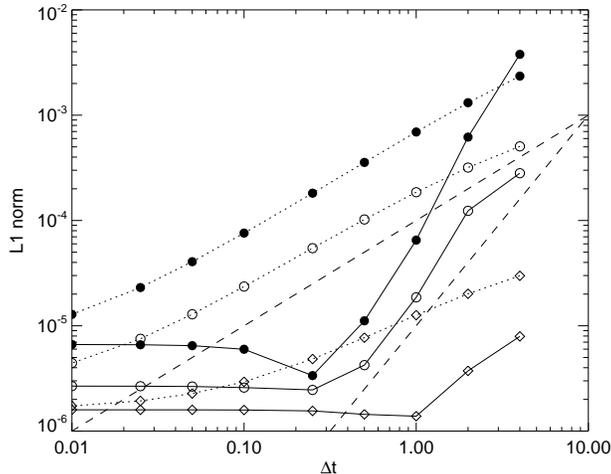}
    \caption{$L_1$ norm of error as a function of time step $\Delta t$
      for the Gaussian diffusion problem.  Solid and dotted lines
      indicate the errors obtained by the schemes with $\lambda=1/2$ and 1,
      respectively.  Diamonds, open circles, and filled circles denote
      the errors on the grids of levels 0, 1, and 2, respectively.  
      The dashed lines indicate
      the relationships of errors in proportion to $\Delta t$ and $\Delta t^2$.
    }
    \label{amrAccuracyFC.eps}
  \end{center}
\end{figure}

Figure~\ref{amrAccuracyFC.eps} shows the $L_1$ norm of the error as a
function of time step $\Delta t$ for $\lambda=1/2$ (solid lines) and 1
(dotted lines).  The norm is estimated separately on each grid level.
The errors for $\lambda=1/2$ and 1 exhibit second-order accuracy and first-order accuracy, respectively, on the grids of levels 1 and 2.  For the
grid of level 0 (the base grid), the dependence of the errors on 
$\Delta t$ is shallower because of the periodic boundary conditions.
Note that the scheme with $\lambda=1/2$ maintains second-order accuracy on the grid of level 2, even when considerable undulation occurs with a large time step.

\subsection{Comparison with an explicit scheme}

We compared the solutions obtained by implicit schemes with the solutions
obtained by an explicit scheme.  The resistivity is distributed as follows:
\begin{equation}
\eta(\mbox{\boldmath$r$}) = \exp\left[-(x^2+ y^2+z^2)\right]
\end{equation}
and the initial magnetic field is given by
\begin{equation}
B_z(\mbox{\boldmath$r$}) = \exp\left[-(x^2+ y^2)\right]
\end{equation}
and $B_x = B_y = 0$.  The computational domain is $x, y, z \in [-4,
4]$, which is resolved by $64^3$ cells. Periodic boundary conditions
are imposed.  We continue the dissipation process of $\mbox{\boldmath$B$}$ until $t
= 1$ by using the implicit schemes with $\lambda=1/2$ and 1 and an
explicit scheme.  The explicit scheme integrates time by means of the
predictor-corrector method and a spatial central difference in order
to achieve second-order accuracy in time and space.  The time step is
set at $\Delta t = 10^{-3}$ for the explicit scheme, and $\Delta t =
10^{-1}$ for the implicit schemes.

\begin{figure}
  \begin{center}
    \FigureFile(40mm,40mm){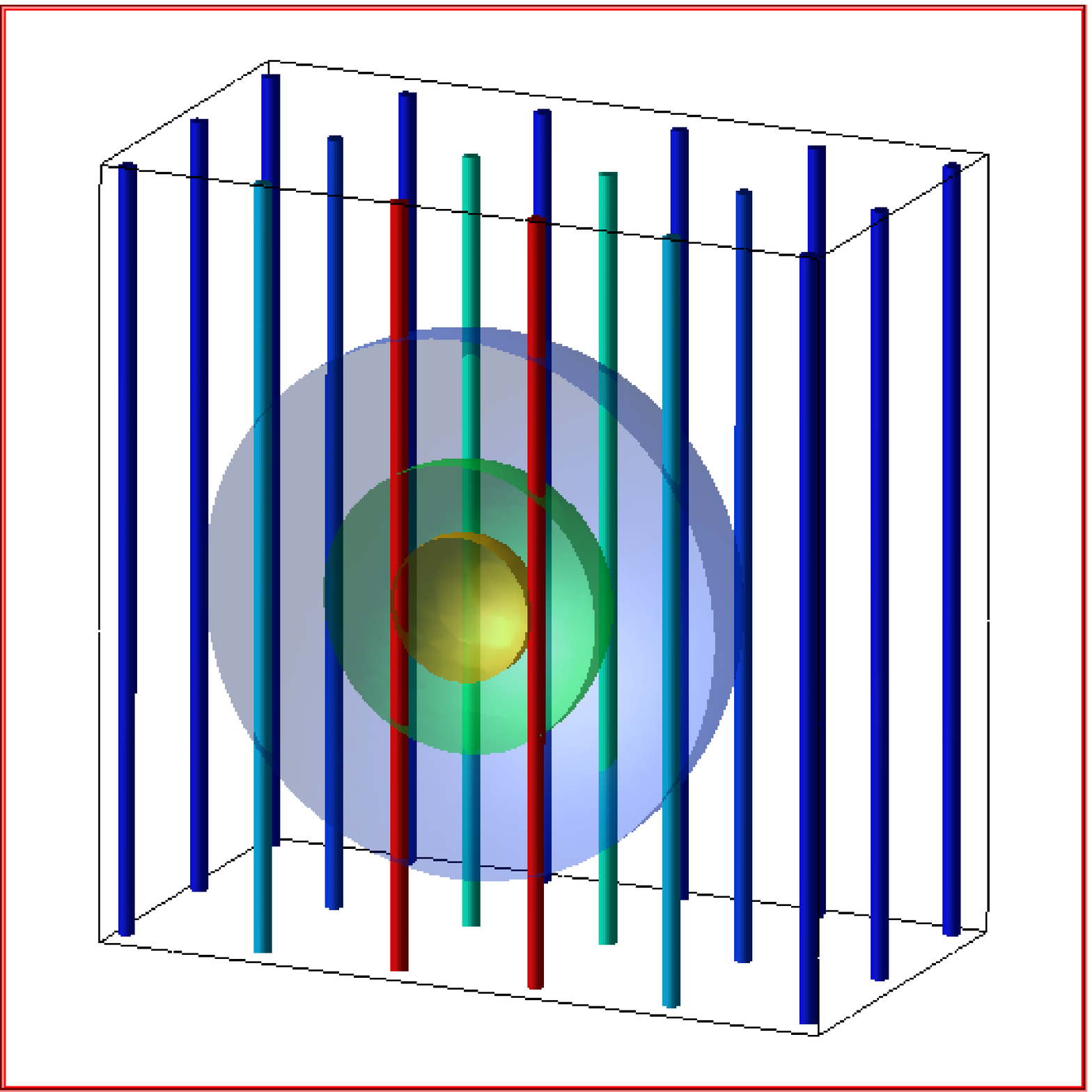}
    \FigureFile(40mm,40mm){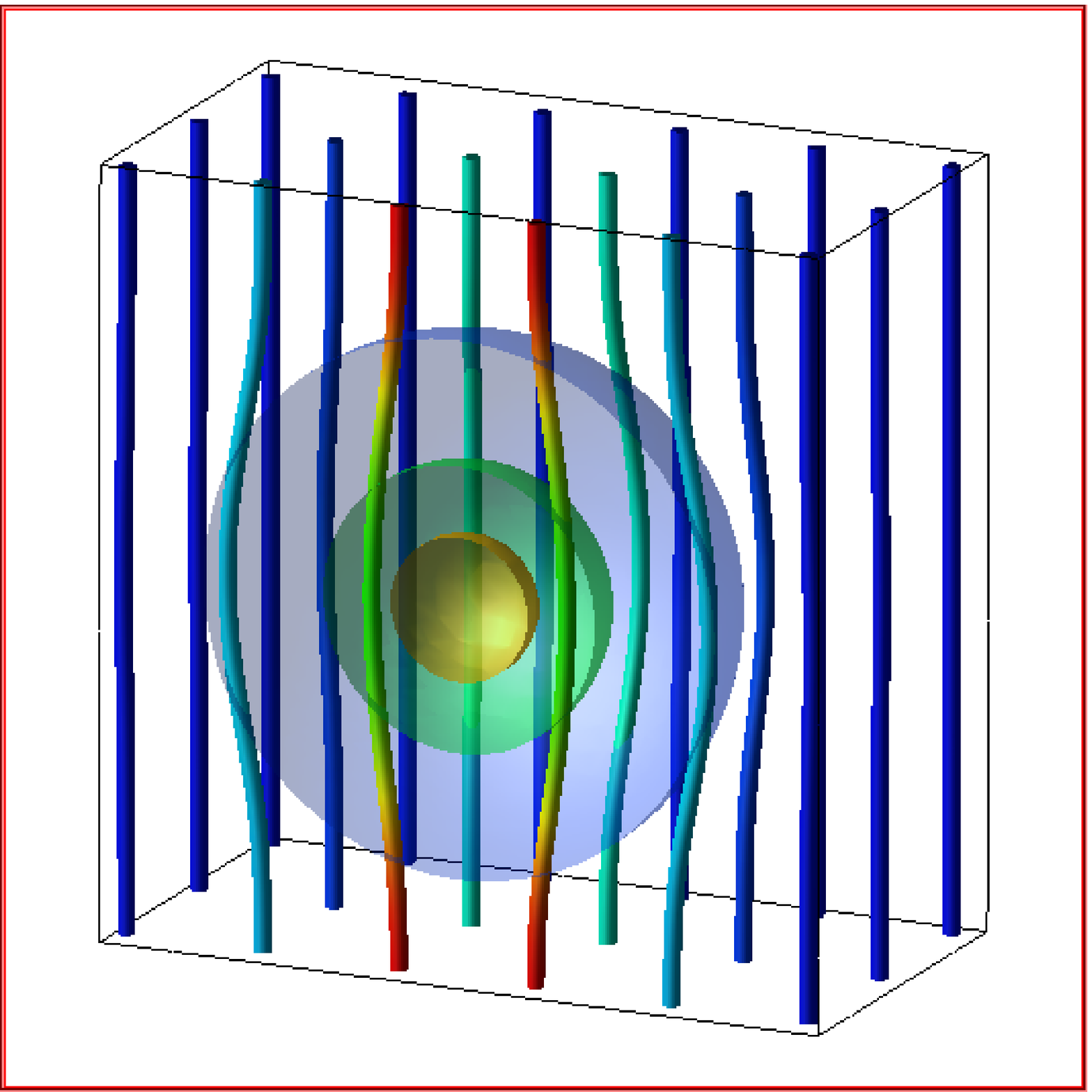}
    \caption{ Distributions of the magnetic field $\mbox{\boldmath$B$}$ and the
      resistivity $\eta$ at the initial stage ({\it left}) and the
      stage of $t = 1$ ({\it right}), which is solved by the second-order implicit scheme with $\lambda=1/2$.  Tubes and isosurfaces
      denote the magnetic field lines and resistivity, respectively.  The colors of
      tubes illustrate the field strength.  The levels of the
      isosurfaces are $\eta = 0.05, 0.425, 0.8$.  The region of $x, z
      \in [-2.4,2.4]$, $y \in [0,2.4]$ is shown.  }
    \label{odTestSphere3d.eps}
  \end{center}
\end{figure}

Figure~\ref{odTestSphere3d.eps} shows the initial conditions and
the solution at $t=1$ solved by the implicit scheme with $\lambda=1/2$.
The right-hand figures show the magnetic fields bent due to the ohmic
dissipation, indicating the reduction of $B_z$ and the generation of $B_x$ and $B_y$.

\begin{figure*}
  \begin{center}
    \FigureFile(55mm,55mm){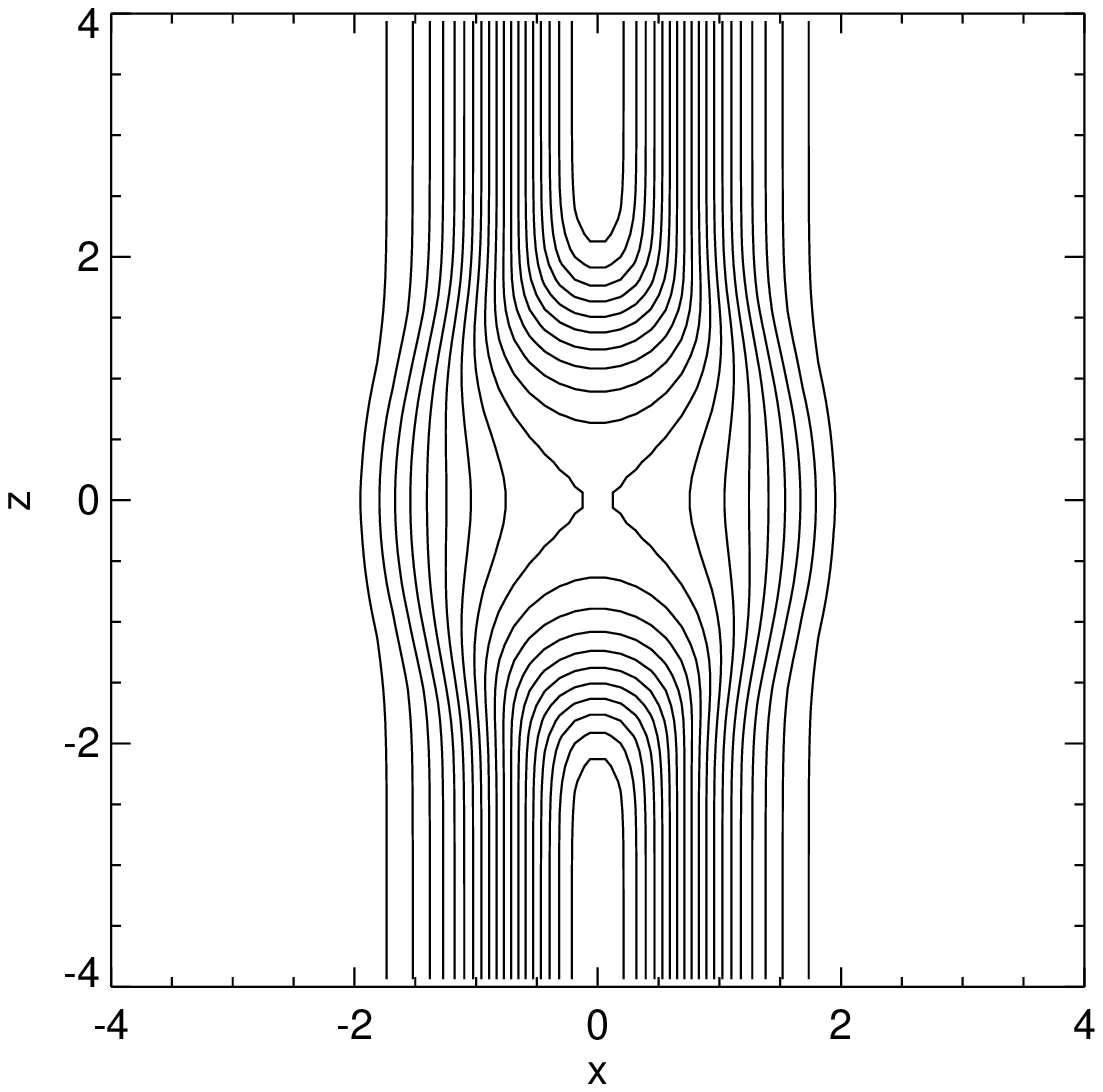}
    \FigureFile(55mm,55mm){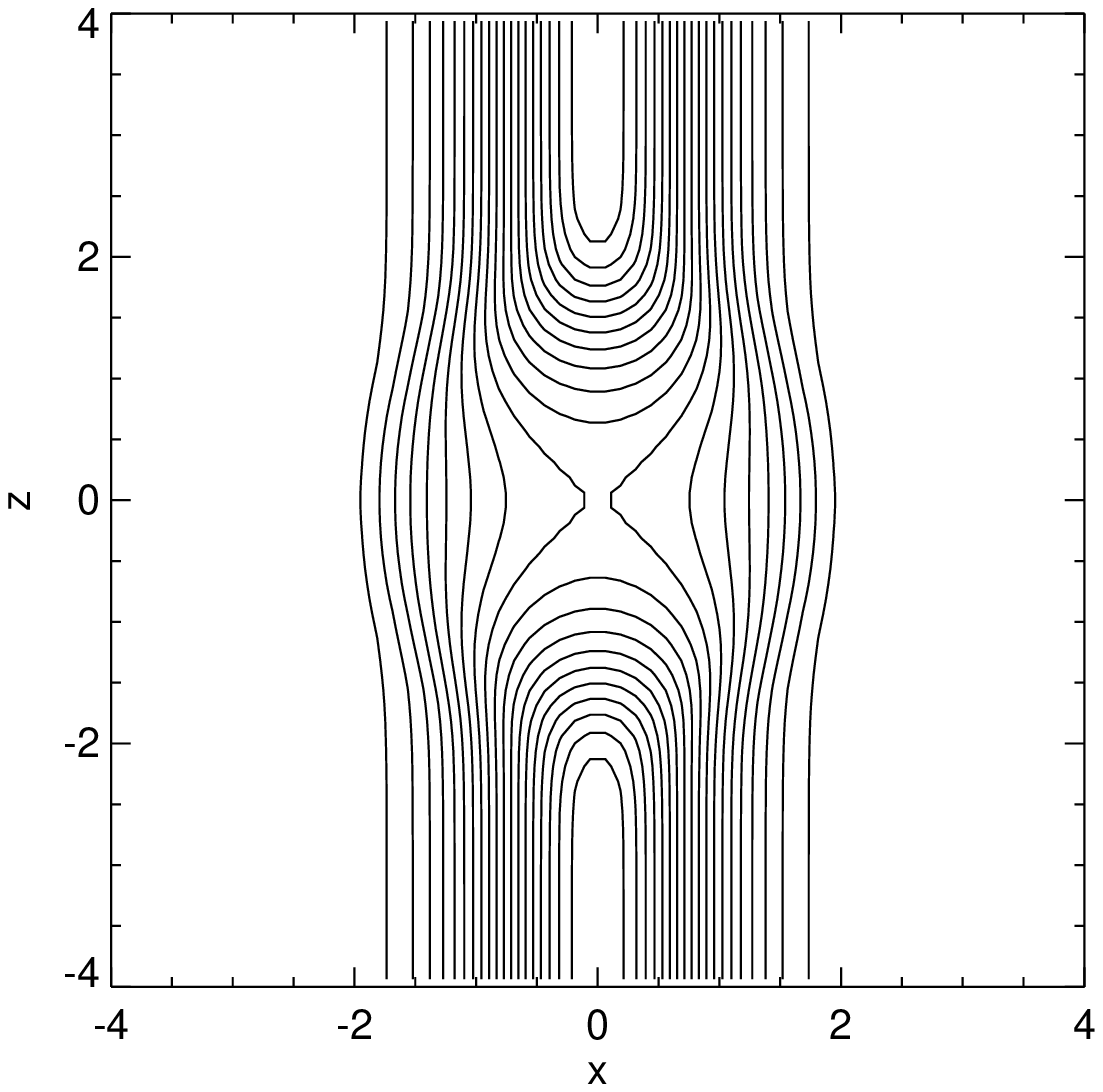}
    \FigureFile(55mm,55mm){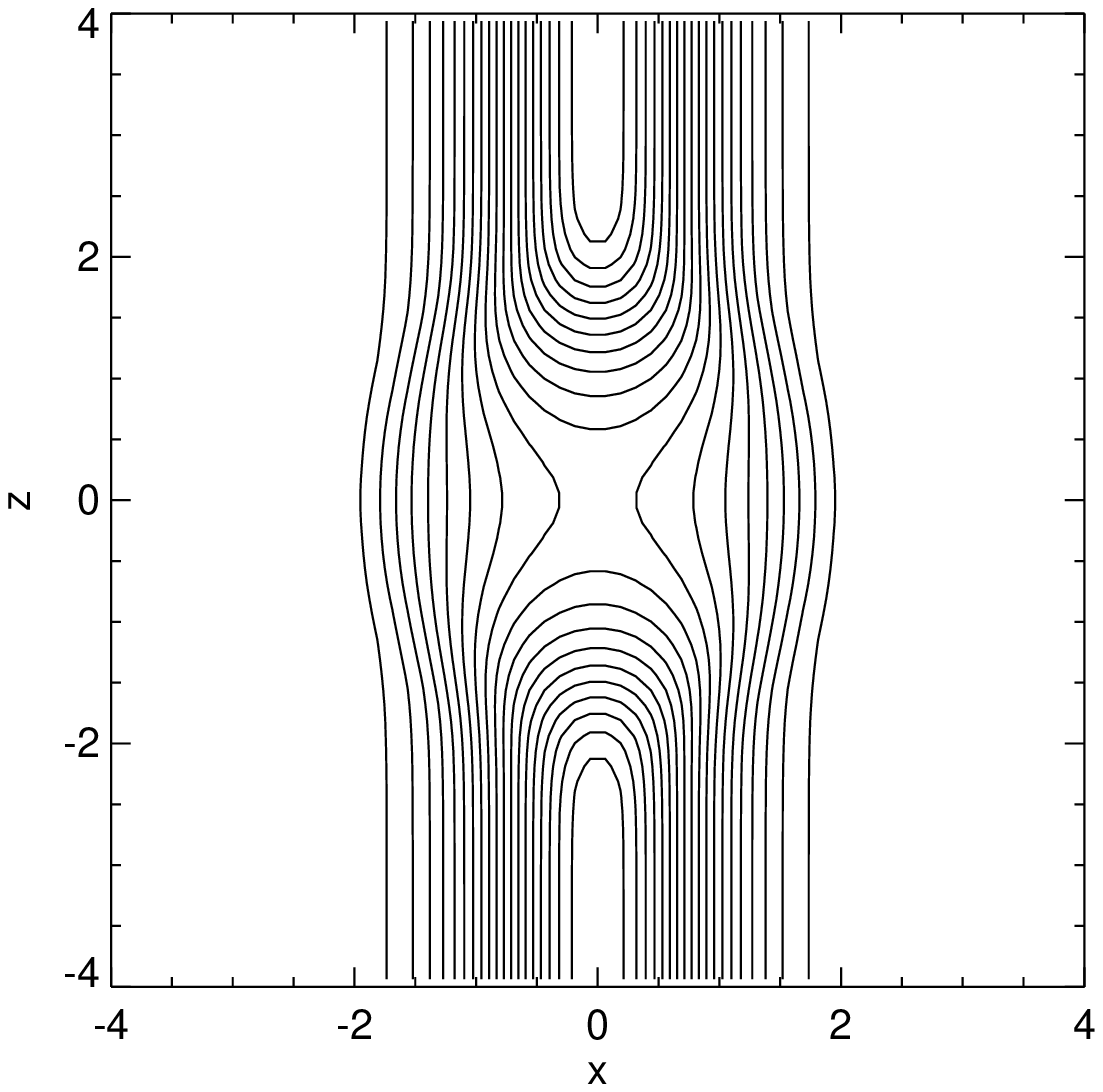}\\
    \FigureFile(55mm,55mm){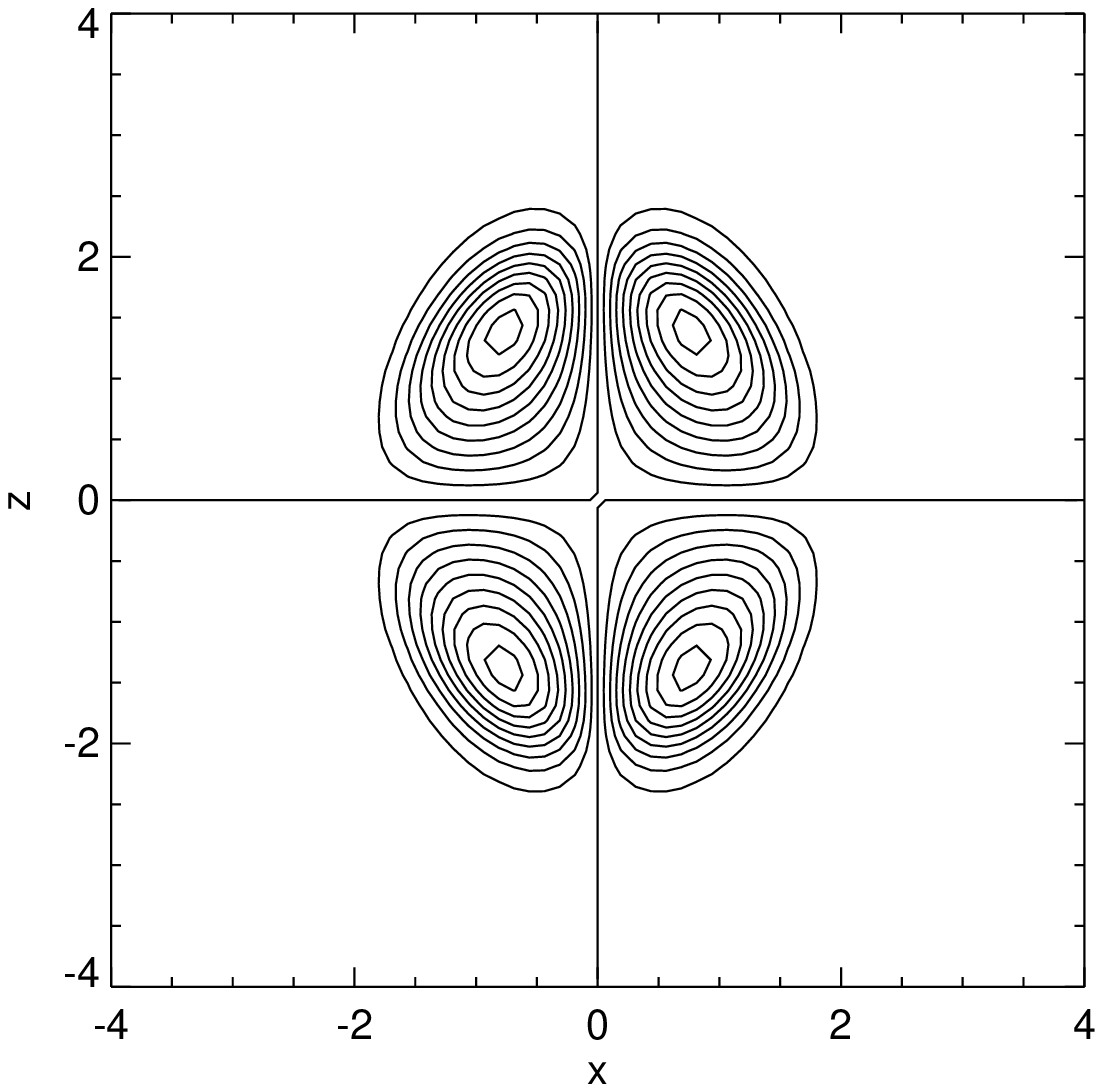}
    \FigureFile(55mm,55mm){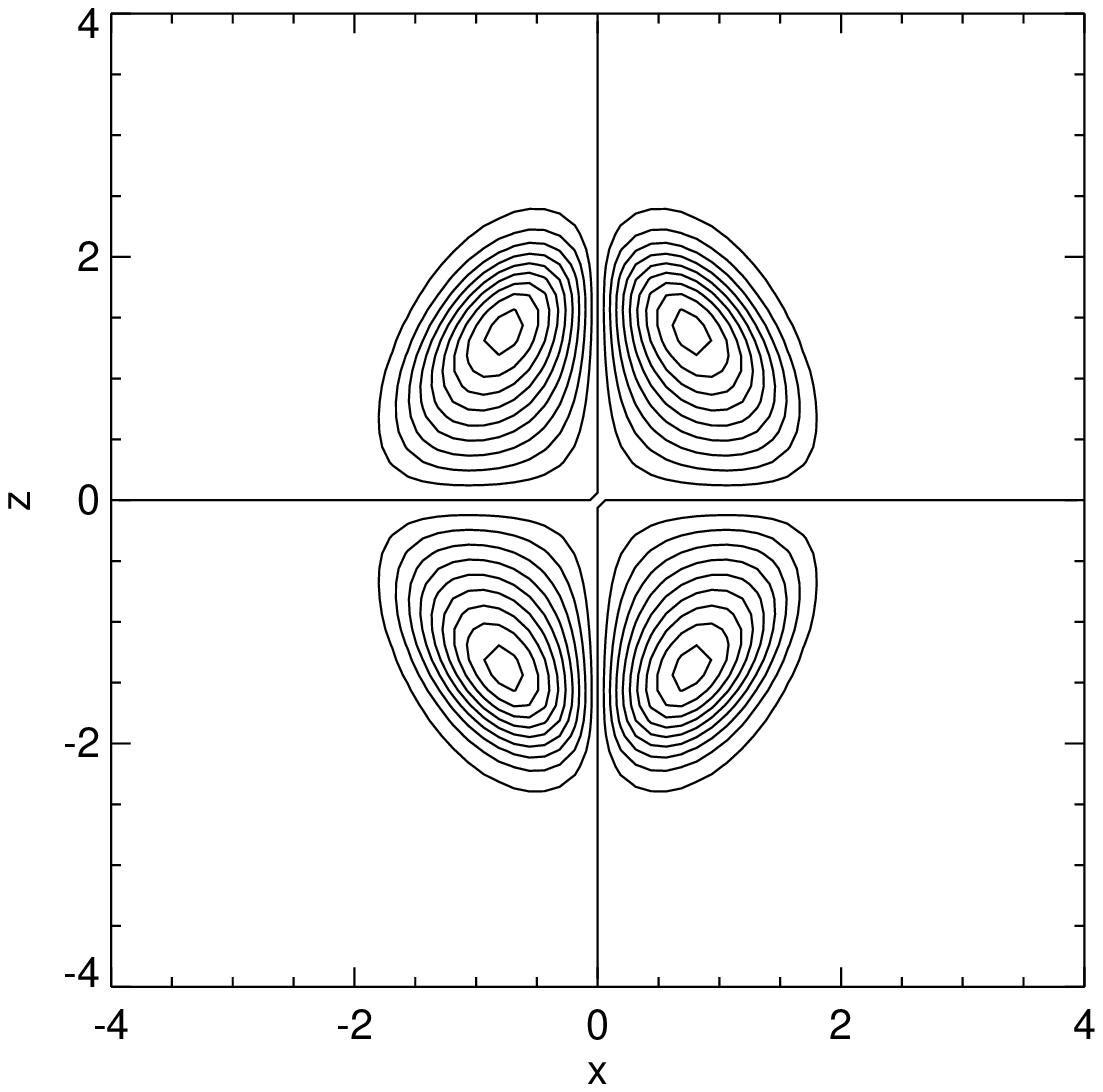}
    \FigureFile(55mm,55mm){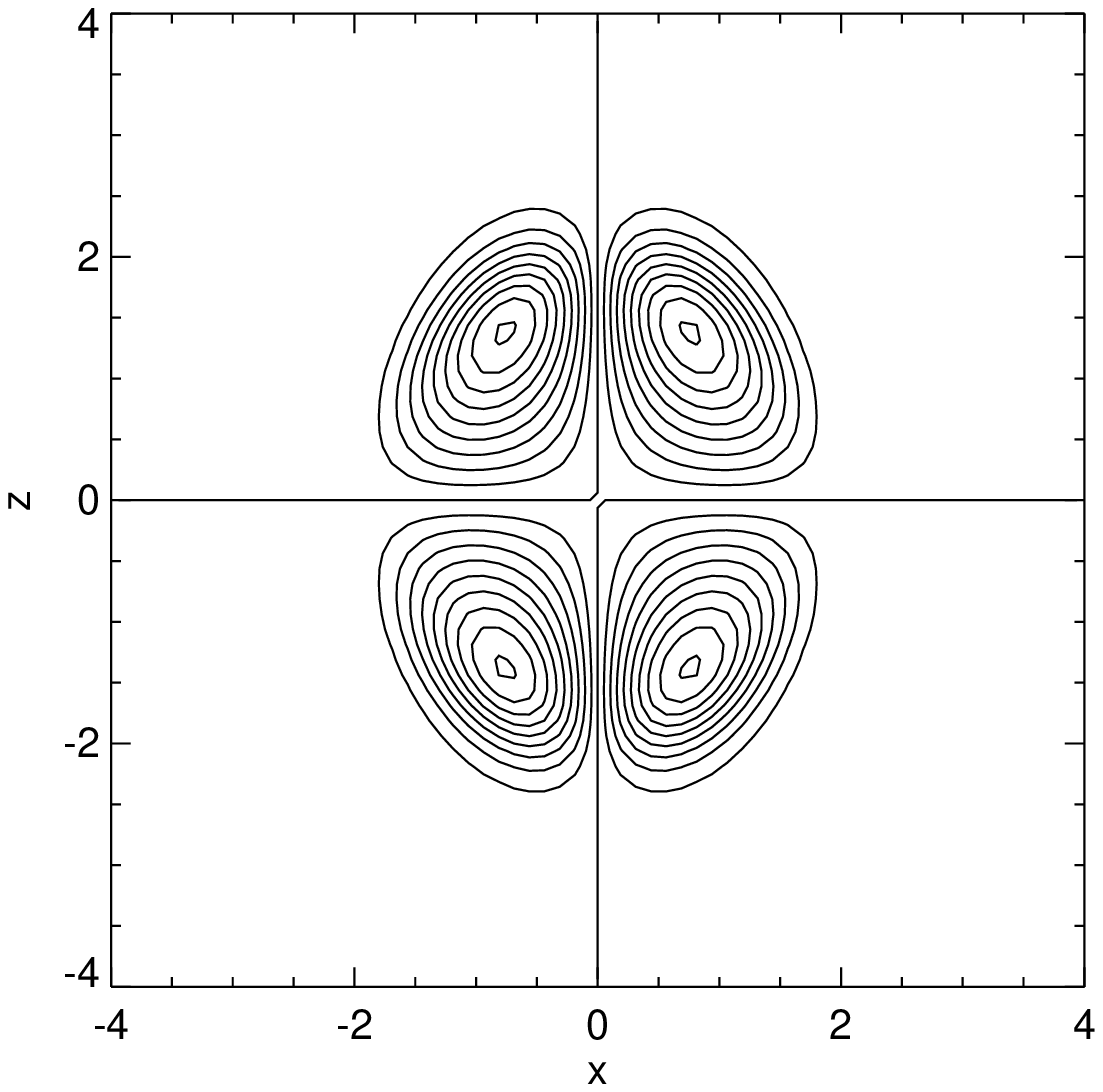}
    \caption{Distributions of the magnetic fields at $t = 1$ in the
      $y=0$ plane. The magnetic fields are solved by 
      the explicit scheme ({\it left}), the implicit scheme with $\lambda=1/2$ ({\it middle}), 
      and the implicit scheme with $\lambda=1$ ({\it right}).
      Upper panels show $B_z$, where
      the contour levels are $B_z = 0.05, 0.1, \cdots, 0.95$.
      Lower panels show $B_x$, where
      the contour levels are $B_x =-0.09, -0.08, \cdots, 0.09 $.}
    \label{odTestSphere.eps}
    \end{center}
\end{figure*}

Figure~\ref{odTestSphere.eps} compares magnetic fields obtained by the
explicit scheme, the second-order implicit scheme ($\lambda=1/2$), and
the first-order implicit scheme ($\lambda=1$).  All of the solutions are
consistent with one another.  In particular, the solution obtained by the
implicit scheme with $\lambda=1/2$ exhibits excellent agreement with
that obtained by the explicit scheme.  This is attributed to the high
accuracy of the implicit scheme with $\lambda=1/2$, which achieves 
second-order accuracy.

\section{Summary and Discussion}

We have presented an implicit scheme for solving the ohmic dissipation for the
SFUMATO MHD-AMR code.  The induction equation of the ohmic
dissipation is solved by this implicit scheme, which is 
based on the Crank--Nicolson method, which has an option for selecting 
first-order accuracy ($\lambda=1$) and second-order accuracy
($\lambda=1/2$) in time.  For both cases, the spatially central
difference yields second order accuracy in space.

The multigrid method is used for the convergence of a solution and exhibits 
fast convergence.  Although the convergence speed depends on $\eta \Delta
t/\Delta x^2$, several cycles of the multigrid method reduce
the residual by more than an order magnitude.
The solution is obtained over the AMR hierarchical grid, in which fine
and coarse grids co-exist. Moreover, no spurious features appear
at the interface between fine and coarse grids.
Note that in the convergence process of the multigrid method, the numerical
fluxes given by equations (\ref{eq:Fx}) through (\ref{eq:Fz}) are
conserved even at the interfaces between the fine and coarse grids,
because of the refluxing procedure, where the fluxes of the coarse grid
are obtained by summing those of the fine grids at the interface
\citep{Matsumoto07}.  This leads to conservation of magnetic flux.

Since the second-order scheme of $\lambda=1/2$ is based on the
Crank--Nicolson method, the scheme is unconditionally stable.
However, the solution can contain spurious oscillations when $\eta
\Delta t/\Delta x^2$ is large, as shown in
Figure~\ref{odGaussianT1_2nd.eps}.  For example, for a one-dimensional
heat equation, the analysis of the maximum principle leads a condition
of $\eta \Delta t/\Delta x^2 \le 3/2$ to suppress the oscillation \citep{Morton05}.
This condition is slightly weaker than the CFL condition of an
explicit scheme.  In the astrophysical simulations, the oscillation of
the magnetic field may change the direction of the
magnetic pressure gradient, which may change the phenomena of the
simulations qualitatively.  In contrast, the first-order scheme
($\lambda=1$) retains monotonicity, as shown in
Figure~\ref{odGaussianT1_1st.eps}, but its error is larger than that
of the second-order scheme.

\bigskip

Numerical computations were carried out on the Cray XT4 at the Center for
Computational Astrophysics, CfCA, of the National Astronomical Observatory
of Japan.
The present research was supported in part by the Hosei Society of
Humanity and Environment. 
The present research was supported in part by
Grants-in-Aid for Scientific Research (C) 20540238
and (B) 22340040 from the Ministry of Education, Culture, Sports, Science and Technology, Japan.

\clearpage

\end{document}